\documentclass[12pt]{article}

\let\a=\alpha

\newcommand{\beq}{\begin{equation}}
\newcommand{\eeq}{\end{equation}}
\newcommand{\beqn}{\begin{eqnarray}}
\newcommand{\eeqn}{\end{eqnarray}}

\newcommand{\eps}{\epsilon}

\newcommand{\be}{\begin{equation}}
\newcommand{\ee}{\end{equation}}
\newcommand{\ba}{\begin{eqnarray}}
\newcommand{\ea}{\end{eqnarray}}
\newcommand{\bdm}{\begin{displaymath}}
\newcommand{\edm}{\end{displaymath}}

\def\a{\alpha}

\newcommand{\ie}{{\it i.e.\ }}
\newcommand{\eg}{{\it e.g.\ }}

\DeclareMathAlphabet{\mathpzc}{OT1}{pzc}{m}{it}
\def\bea{\begin{eqnarray}}
\def\eea{\end{eqnarray}}
\def\beas{\begin{eqnarray*}}
\def\eeas{\end{eqnarray*}}
\def\sla{\raise.15ex\hbox{$/$}\kern-.57em}

\def\bea{\begin{eqnarray}}
\def\eea{\end{eqnarray}}
\def\de{\partial}

\def\sla{\raise.15ex\hbox{$/$}\kern-.57em}
\def\ie{{\it i.e.}~}
\def\eg{{\it e.g.}~}
\def\ap{{\alpha^\prime}}

\def\a{\alpha}

\def\cA{{\cal A}}
\def\cB{{\cal B}}

\def\cE{{\cal E}}
\def\cF{{\cal F}}
\def\cG{{\cal G}}
\def\cH{{\cal H}}
\def\cI{{\cal I}}

\def\cK{{\cal K}}
\def\cL{{\cal L}}
\def\cM{{\cal M}}
\def\cN{{\cal N}}

\def\cP{{\cal P}}
\def\cQ{{\cal Q}}
\def\cR{{\cal R}}
\def\cS{{\cal S}}
\def\cT{{\cal T}}
\def\cU{{\cal U}}
\def\cV{{\cal V}}
\def\cW{{\cal W}}

\begin{document}
%
%
\begin{titlepage}
\vskip 2cm
\begin{center}
{\Large\bf Recent Trends in Superstring
Phenomenology\footnote{Lectures delivered at {\it Foundations of
Space and Time - Reflections on Quantum Gravity} STIAS,
Stellenbosch, South Africa, 10-14 August 2009}}
\end{center}
\vskip 2cm
\begin{center}
{\large\bf Massimo Bianchi}~\\
{\sl Dipartimento di Fisica and Sezione I.N.F.N. \\ Universit\`a di Roma ``Tor Vergata''\\
Via della Ricerca Scientifica, 00133 Roma, Italy}\\

\end{center}
\vskip 3.0cm
\begin{center}
{\large \bf Abstract} \end{center}

 We review for non-experts
possible phenomenological scenari  in String Theory. In particular
we focus on vacuum configurations with intersecting and/or
magnetized unoriented D-branes. We will show how a TeV scale
tension may be compatible with the existence of Large Extra
Dimensions and how anomalous U(1)'s can give rise to interesting
signatures at LHC or in cosmic rays. Finally, we discuss
unoriented D-brane instantons as a source of non-perturbative
effects that can contribute to moduli stabilization and susy
braking in combination with fluxes. We conclude with an outlook
and directions for future work.



\vfill

\end{titlepage}


\section*{Foreword}

After more than 40 years, there is no experimental evidence for
String Theory or, else, a satisfactory -- possibly holographic --
description of the QCD string is not yet available.

Still, the Veneziano model predicts a massless vector boson in the
open string spectrum and the Shapiro-Virasoro model a massless
tensor boson in the closed string spectrum. These two particles
can be naturally associated to the two best known forces in
Nature: gravity and electro-magnetism.

Moreover, String Theory makes a definite -- albeit incorrect --
prediction for the number of Space-Time dimensions: 26 for Bosonic
Strings, 10 for Superstrings. This basic fact led to many so-far
unsuccessful attempts to compactifications: Calabi-Yau, orbifolds,
non-geometric models, ...

Only relatively recently, it has been fully appreciated that one
needs fluxes and non-perturbative effects for stabilizing moduli,
breaking susy, ... and eventually making contact with
astro-particle phenomenology.

The plan of this lecture is to first briefly present the available
phenomenological {\it scenari}. Starting from Heterotic Strings on
CY 3-folds and un-oriented Strings (Type I and other Type II {\it
un}-Orientifolds), we will eventually discuss F-theory on elliptic
CY 4-folds and M-theory on singular $G_2$-holonomy spaces and
consider the role of fluxes, dualities and (Euclidean) branes in
the characterization of an interesting portion of the Landscape of
vacua.

We will then focus on the case that most easily allows the
embedding of the (N-MS)SM (Not-necessarily Minimal Supersymmetric
Standard Model) in String Theory \ie intersecting and/or
magnetized unoriented D-branes. We will show how a rather Low
String Tension (TeV scale) is compatible with the existence of
Large Extra Dimensions and how anomalous $U(1)$'s could give rise
to spectacular signatures at LHC or in cosmic rays. Finally, we
will discuss unoriented D-brane instantons as a source of
non-perturbative superpotentials and other effects that crucially
contribute to moduli stabilization and susy braking in combination
with fluxes.

We will conclude with an outlook and directions for near future
work.

\section*{Green-Schwarz mechanism and Heterotic Strings}

25 years ago\footnote{ {\it Anomaly cancellations in
supersymmetric D = 10 gauge theory and superstring theory}, by
Michael B. Green and John H. Schwarz, Physics Letters B Volume
149, Issues 1-3, 13 December 1984, Pages 117-122 Received 10
September 1984.} Michael Green and John Schwarz proposed a
mechanism for anomaly
  cancellation in Type I superstrings (unoriented open and closed strings)
  with $SO(32)$ gauge group.
 Assigning non-linear gauge transformation properties to
  the Ramond-Ramond (R-R) 2-form
$$ \delta_{_{YM}} C_{\mu\nu} = Tr( \alpha_{_{YM}} F_{\mu\nu}) + ... $$
and including a local counterterm of the schematic form $$ L_{GS}
= C \wedge Tr(F\wedge F \wedge F \wedge F) + ... $$ in the
effective action restores gauge invariance at the quantum level.
The abstract ended with the statement ``... A superstring theory
for $E(8)\times E(8)$ has not yet been constructed.'' The answer
to the problem is the {\it Heterotic String}\footnote{ {\it
Heterotic String} by David J. Gross, Jeffrey A. Harvey, Emil
Martinec, and Ryan Rohm Phys. Rev. Lett. 54, 502 - 505 (1985)
Received 21 November 1984} that combines the left-moving
superstrings with right-moving
  bosonic strings. The difference 26-10 =16 accounts for the rank of
  either $SO(32)$ or $E(8)\times E(8)$.
Compactifications on Calabi-Yau three-folds were soon recognized
as a very promising framework to produce chiral $\cN=1$
supersymmetric models with $E(6)$ gauge group, via `standard
embedding' of the spin connection in the gauge group, and $N_{gen}
= \chi(CY)/2$  \cite{Candetal}. For roughly ten years, from 1985
to 1995, Heterotic Model Building attracted the attention of the
vast majority of string theorists in the hope that non-standard
embeddings (`stable
  holomorphic vector bundles'), giving rise to $SO(10), SU(5), ...,
  SU(3)\times SU(2) \times U(1)$ gauge groups and
  $N_{gen} \neq \chi (CY)/2$, could produce acceptable particle
  phenomenology\cite{HetPhen}.

Many problems however were left open \cite{Kirbook}
\begin{itemize}
  \item{Too many {\it moduli} (massless scalars / flat directions)}
    \item{Hard to get precisely $N_{gen} = 3$ and no vector-like pairs}
  \item{Hard to avoid chiral exotics and/or fractional charges}
\item{Discrepancy between $M_{Pl} \approx {M_s \sqrt{V_{int}}\over
g_s}$ and $M_{GUT}$ since $\alpha_{GUT} \approx {g_s^2\over
V_{int}} \approx {1\over 25}$} barring large threshold corrections
  \item{No viable SUSY breaking without tachyons and small dark energy density}
  \end{itemize}

Still, there were important achievements \cite{Kirbook}

\begin{itemize}
\item{World-sheet conformal invariance and moduli spaces of Riemann surfaces}
\item{On-shell Low-energy Effective Supergravity Actions from
scattering amplitudes} \item{T-duality: windings $\leftrightarrow$
KK momenta, Neumann b.c. $\leftrightarrow$ Dirichlet b.c.}
\item{Mirror Symmetry and Quantum/Stringy Geometry}
\item{Non-geometric models (orbifolds, free fermions, Gepner,
...)}
\end{itemize}

\section*{Type I and other Type II {\it un-Orientifolds}}

As already mentioned, theories with unoriented open and closed strings
have the virtue of predicting gravitational interactions mediated by
closed strings and gauge interactions mediated by open strings.

In 1987, Augusto Sagnotti proposed to view the Type I superstring
in $D=10$ as {\it parameter space orbifold} or {\it open string
descendant} of the Type IIB superstring \cite{Aug}

For almost ten years, from 1986 to 1996, Type I model building
{\it `without'} direct mention of D-branes and $\Omega$-planes was
systematically developed by the group in Tor Vergata
\cite{CAASrev}. The results were rather surprising
\begin{itemize}
\item{Non-supersymmetric (non) tachyonic models \cite{MBASGP}}
\item{Vertex operators for R-R potentials \cite{BPSt}} \item{Rank
reduction with $B_{NS-NS}\neq 0$ \cite{BPSt}} \item{Several tensor
mltps in $D=6$ \cite{MBAS123}, leading to a generalized GS
mechanism \cite{GSSmech}} \item{$\cN=1$ susy chiral models in
$D=4$ with $N_{gen}=3$ \cite{ABPSS}} \item{Magnetic fluxes and
SUSY breaking \cite{MagnDb}} \item{Boundary Conformal Field Theory
\cite{BCFT}}
\end{itemize}

It is fair to say, however, that the geometric picture  whereby
open strings terminate on D-branes \cite{DaiLP} which carry R-R
charge \cite{JoeP} was crucial in all later development including
holography.

\section*{Branes and dualities}

Polchinski's observation was not isolated. In addition to
fundamental strings (`one-branes') String Theory indeed admits
other 1/2 BPS ($T_p = |Q_p|$) extended solitonic solutions, called
p-branes \cite{HullTownDemo}, that couple to different higher rank
antisymmetric tensor fields , \ie $(p+1)-$forms such as
$B_{\mu\nu}^{(2) NS-NS}$ and $C^{(p+1)
R-R}_{\mu_1\mu_2...\mu_{p+1}}$
\begin{itemize}

\item{NS5-brane couples to $\tilde{B}_6$ which is  the magnetic
dual to ${B}_2$ (sourced by F1) \cite{NS5}}

\item{D-branes are hypersurfaces where open strings can end
\cite{JoeP}

\begin{itemize}

\item{D0-, D2-, D4-, D6-, D8-branes are BPS in Type IIA}

\item{D(-1)-, D1-, D3-, D5-, D7-, D9-branes are BPS in Type IIB}
\end{itemize}}
\end{itemize}

 The generalized `tension' \ie energy per unit world-volume is a distinctive
 feature of each kind of branes: for fundamental strings
 $T_{F1}\approx 1/\ap$, for NS5-branes
 $T_{NS5}\approx 1/g_s^2(\ap)^3$, for Dp-branes $T_{Dp}\approx
 1/g_s(\ap)^{p+1\over 2}$.

Moreover, although one knows very well the micro-scopic
world-sheet  description of a single fundamental string and, to a
lesser extent, of a single NS5-brane. It is not clear how to
describe the dynamics of a stack of strings let alone for
NS5-branes. On the contrary, at least at  low-energy, the dynamics
of a stack of $N$ Dp-branes is known to be governed by the
dimensional reduction to $D=p+1$ of $D=10$ $\cN = (1,0)$ SYM in
$D=10$ with $U(N)$ gauge group \cite{WitDb}.

The list of extended objects includes the rather peculiar
Orientifold planes \cite{DaiLP}, or simply $\Omega$-planes,
(perturbatively) non dynamical BPS solitons with negative tension,
which can neutralize the R-R charge of D-branes. $\Omega$-planes
act as `mirrors' both on the target space and on the world-sheet:
$X_{\perp} (z, \bar{z}) \rightarrow - X_{\perp} (\bar{z},z)$ and
allow D-branes with $SO(N)$ and $Sp(2N)$ gauge groups or
(anti)symmetric tensor representations of $U(N)$.

\section*{String duality conjectures}

As a result of  the existence of BPS p-branes, which may become
light under suitable conditions, different string theories may
offer complementary descriptions of the same Physics in different
regimes \cite{HullTownU,WitDynVD}.

The best tested duality conjectures are:

\begin{itemize}

\item{Type IIA at strong coupling exposes one more dimension  and
gives rise to Supergravity in $D=11$ at low-energy
\cite{HullTownU,WitDynVD}. The dilaton is related to the diagonal
component of the metric in the extra dimension $\phi \approx
G_{11,11}$, while the R-R vector and the NS-NS 2-form arise from
the mixed components of the metric and 3-form, respectively.
D0-branes (D-particles) are nothing but KK excitations while all
other branes can be related to M2-branes, sourced by the 3-form,
their magnetic dual M5-branes in $D=11$ and purely gravitational
solitons. }

\item{$SO(32)$ Heterotic and Type I strings are S-dual in $D=10$
\cite{PolWit,HullHet,Dabhet}. The effective actions coincide up to
field redefinitions \be \phi_H = - \phi_I \quad G_{MN}^H =
e^{-\phi_I} G_{MN}^I \ee The $SO(32)$ heterotic string can be
identified with the Type I D-string since $T_H = T_I/g_s^I$, while
Type I fundamental strings being unoriented are not BPS
objects\footnote{They are  `sourced' by the NS-NS 2-form which is
projected out by $\Omega$.}. The duality entails strong
$\leftrightarrow$ weak coupling exchange, \ie $g_s^I = 1/g_s^H$.
 The duality has not only been accurately tested at the level of BPS
 saturated couplings in toroidal and
some orbifold compactifications \cite{IHtests} but also at the
level of some massive (non) BPS states such as the 128 spinorial
heterotic states that correspond to a Type I D-particle in $D=10$
\cite{nonBPS}. }

\item{Heterotic on $T^4$ is dual to Type IIA on $K3$
\cite{HetIIAK3}. The heterotic string can also be viewed as the
Type IIA NS5-branes wrapped around a $K3$ surface and {\it  vice
versa}.} The duality has been accurately tested at the level of
BPS saturated couplings in $D=6$ and in toroidal
compactifications.

\end{itemize}

The picture that emerges is that all known superstring theories
can be related to one another by the action of some discrete
transformation of the so-called U-duality group \cite{HullTownU}.
The theory that subtends the network has been (provisionally)
called M-theory \cite{WitDynVD}. U-duality transformations are the
discrete remnants of the non-compact symmetries of extended
supergravities. The existence of (discrete) U-duality
transformations acting non-linearly on massless scalar fields
opens the door to new kinds of compactifications or rather vacuum
configurations whereby scalars are non-trivially identified under
monodromies. The most prominent one, as we will see later, goes
under the name of F-theory and exploits the SL(2) symmetry of Type
IIB superstrings acting non-linearly on the complexified dilaton
$\tau = \chi + i e^{-\phi}$ \cite{VafaFt}.  $\tau$  can be
considered as the complex structure modulus of an auxiliary torus
(elliptic curve) that is non-trivially fibred over `physical'
space-time.

The study of brane dynamics led Juan Maldacena to conjecture
another quite remarkable duality \ie the holographic
correspondence between Type IIB in anti De Sitter space and
$\cN=4$ SYM on the boundary \cite{JMholoetal}. For the first time
the long-sought for string description of large N gauge theories
was made concrete albeit for a very special gauge theory that is
superconformal and does not `confine' in the usual sense.
Moreover, the holographic principle for gravity of 't Hooft and
Susskind finds a precise realization. Maldacena correspondence
proved to be a gold mine. In addition to being tested and
generalized to many other forseeable contexts, including warped
extra-dimensional scenari \cite{RS12}, it has lead to
integrability and to unforseeable applications to non-relativistic
condensed matter and astrophysical systems \cite{HoloPhen}.

\section*{Phenomenological scenari}

Two of the key features of the Standard Model are the chiral
spectrum of elementary fermions and the existence of a light
scalar field, the Higgs boson. In order to accommodate chirality
and protect the hierarchy between the weak scale (1 TeV) and any
fundamental scale ($10^{16-19} GeV$) vacuum configurations with
(softly broken) $\cN=1$ SUSY to $D=4$ are definitely the best
candidate. In String Theory, this leads to the following options

\begin{itemize}

\item{Heterotic strings on CY 3-folds with non-trivial vacuum gauge
bundles}

\item{Intersecting / magnetized D-branes in Type II {\it
un-}Orientifolds}

\item{F-theory on elliptic CY 4-folds}

\item{M-theory on singular $G_2$ holonomy spaces}

\end{itemize}

In order to stabilize moduli and break susy one needs to turn on
internal fluxes \cite{GKP}, \ie gauging some of the continuous
symmetries in the supergravity description \cite{FluxGaugSUGRA},
and include non-perturbative effects from Euclidean p-branes
wrapping internal cycles, \ie `p-brane instantons'
\cite{unDbinst}.

\section*{Heterotic on CY 3-folds with fluxes / branes}

For historical reasons, the first class of  phenomenologically
viable models correspond to heterotic on CY 3-folds. Before
including fluxes and (non)perturbative effects, chosen a suitable
CY 3-fold, one has to find a non-trivial vacuum gauge bundles  to
produce some susy extension of the standard model.

One has to heavily rely on the algebro-geometric construction of
stable holomorphic vector bundles \cite{GSW}. The choice of the
structure group determines the resulting GUT group: $SU(3), SU(4),
SU(5)$ respectively correspond to $E(6), SO(10), SU(5)$ GUT's from
the visible $E(8)$. The hidden E(8) gives rise to gaugino
condensation that can produce a field-dependent superpotential $W
= \Lambda^3_{SYM}(\phi_i)$. In addition to the obvious dependence
on the 4-d dilaton, gauge couplings indeed acquire one-loop
threshold corrections that depend on the compactification moduli.
This simple $W$ leads to a runaway potential. A variant with
several gaugino condensates at different scales, known as
`race-track' models, seems more appealing \cite{racetrack}. Yet
the inclusion of 3-form and metric (torsion) fluxes as well as
branes seems unavoidable \cite{HetTors}.

NS-NS fluxes are amenable to a world-sheet analysis and may even
lead to interesting non-geometric constructions involving
T-duality. T-folds correspond to gaugings of discrete non-compact
symmetries which do not correspond to geometric manifolds, in that
the gluing of the various patches requires identifications which
are not simply diffeomorphisms. This is only possible for theories
with stable extended solitons (p-branes) and goes beyond the
point-particle description. A trick that allows to deal with some
non-geometric construction in more geometric terms at least for
tori is the `doubling' of all internal coordinates
\cite{HullDoubling}.

It is however an unsolved problem how to include  (wrapped) NS5's
even in geometric backgrounds, let alone in non-geometric
configurations. On the contrary, D-branes admit a CFT description,
that will prove extremely useful in the sequel, and can be easily
studied in non-geometric backgrounds \cite{MBLRasymmDb}.

\section*{Un-oriented D-brane worlds}

As pioneered by  Augusto Sagnotti and the Tor Vergata group
\cite{CAASrev}, vacuum configurations with open and closed
unoriented strings can be thought of `descendants' of Left-Right
symmetric theories with oriented strings. In more geometric terms,
this means including D-branes and, for consistency,
$\Omega$-planes \cite{Uberalles}.

Depending on the choice of parent theory  and $\Omega$ projection
one has the following options:

\begin{itemize}

\item{Type I \ie Type IIB with  $\Omega$9-plane wrapping the
internal space and implementing a $Z_2$ projection of the closed
string spectrum under worldsheet parity $\Omega$. Additional
$\Omega$5-planes wrapping (collapsed) 2-cycles may be present. One
can then include (magnetized) D9-branes and D5-branes wrapping
holomorphic 2-cycles with $F_3$ fluxes}

\item{Type IIB with $\Omega$3-planes  localized at points in the
internal space and implementing a $Z_2$ projection of the closed
string spectrum $\Omega_B = \Omega (-)^{F_L}\cI$ where $\Omega$
denotes world-sheet parity and $\cI$ is an holomorphic involution
(locally $z^I \leftrightarrow - {z}^I$) whose fixed-point set is a
collection of points and, possibly, 4-cycles, the latter leading
to $\Omega$7-planes.  One can then include D3-branes and
(magnetized) D7-branes wrapping holomorphic 4-cycles with $F_3 +
\tau H_3 = G_3$ fluxes}

 \item {Type IIA with
$\Omega$6-planes wrapping a  3-cycle and implementing a $Z_2$
projection of the closed string spectrum $\Omega_A = \Omega \cI$
where $\Omega$ denotes world-sheet parity and $\cI$ is an
anti-holomorphic involution (locally $z^I \leftrightarrow -
\bar{z}^I$) whose fixed-point set is a collection of
$\Omega$6-planes.  One can then include intersecting D6-branes
wrapping SLAG's (Special Lagrangian 3-cycles) with various $F_0,
F_2, F_4$ R-R fluxes as well as some $H_3$ and metric fluxes}

\end{itemize}

In any case, $\Omega$-planes with negative tension and R-R charge
are needed for tadpole cancellation (Gauss law) \ie anomaly
cancellation (G-S strikes back!) \cite{MBJFM,AIUM} \eg
$$ N_{D3} + \int_{\cM_6} H_3 \wedge F_3 + \int_{\cW^{(7)}_4}
Tr(F_2 \wedge F_2) = {1\over 4} N_{{\Omega}3} $$ In some cases,
`orientifold' singularities can be resolved  (non) perturbatively
as for instance in F-theory, whereby $\Omega$7-planes are replaced
by collections of mutually non local 7-branes \cite{SenFt}. Since
T-duality or appropriate generalizations, such as mirror symmetry,
should relate the above {\it un-}orientifolds to one another, the
very notion of $\Omega$-planes which is intrinsically perturbative
(and quite counter-intuitive in-so-far as negative tension is
concerned for instance) should be replaced by more sophisticated
constructions generalizing F-theory by the extension of $SL(2)$ to
larger monodromy groups. In the same vein, geometric and
non-geometric fluxes should be unified in a common framework and
associated to different choices of embedding tensor that encodes
the allowed gaugings in the supergravity description
\cite{FluxGaugSUGRA}.

From the world-sheet  vantage point, it is a long-standing problem
to quantize strings in the presence of R-R fluxes, even in
extremely symmetric cases such as $AdS_5\times S^5$. Some progress
can be made resorting to the pure spinor formalism \cite{PureSpin}
or other hybrids \cite{Hybrid} ... but the path is long.  Except
for very special cases such as `deformed' (super) WZW models
\cite{superWZW} or some plane waves \cite{BMNetal}, not even the
free string spectrum is known, let alone the interactions. From a
different perspective, the recent discovery that $\cN =4$ SYM
theory may be integrable in the planar limit has opened the way to
significant progress in the calculation of anomalous dimensions of
composite operators holographically dual to masses of string
states in $AdS$ with R-R flux \cite{BeisRev}. Although, explicit
computations even for the simplest multiplet, known as the Konishi
multiplet \cite{KonNoi}, may require days of computer time
\cite{Vieiraetal}, one should try and be optimistic!

\section*{M-theory phenomenology}

Type IIA at strong coupling grows another dimension
\cite{WitDynVD}: $\cM \rightarrow \cM' \approx \cM \times S^1$.
The dilaton, parameterizing the extra circle, is fibred over the
base. Non trivial fibrations correspond to intrinsically
non-perturbative constructions (in $g_s$). In particular
compactifications on CY 3-folds are replaced by compactifications
on 7-manifolds. In order to preserve $\cN =1$ SUSY in $D=4$ (4 out
of the  original 32 supercharges) the holonomy should be $G_2$
\cite{AchaWit}. Indeed under $SO(7) \rightarrow G_2$ the spinor
decomposes according to ${\bf 8} \rightarrow {\bf 1} + {\bf 7}$.
The singlet corresponds to unborken susy. If the manifold is
smooth, however, no chirality is generated. One has to consider
M-theory on singular $G_2$ holonomy spaces. Very much as in CY
3-folds, manifolds with $G_2$ holonomy admit a globally defined
3-form, $\Phi_3$ known as the associative 3-form. The
co-associative 4-form $*\Phi_3 = \tilde{\Phi}_4$, represents a
natural choice of $G_4$ flux. In the G(2)-MSSM framework proposed
by \cite{AchaKaneKumar, ...}, however, no $G_4$ flux is needed.
Built-in non-perturbative effects such as `racetrack'-like gaugino
condensates \cite{racetrack} produce a superpotential that can
stabilize all moduli and produce peculiar (N-MS)SM spectra. The
explicit construction of the  singular manifold of $G_2$ holonomy
with the desired properties, such as the appropriate singularity
structure and a small number of moduli to start with, is an open
issue.

At any rate, despite the inclusion of a class of non-perturbative
effects, the M-theory framework  is only computationally reliable
in the low-energy supergravity approximation since a truly
microscopic description is still lacking. Only very recently,
significant progress has been made in the understanding of the
dynamics of stacks of M2-branes \cite{M2new}. Very little is known
about coincident M5-branes.

\section*{F-theory on elliptic CY 4-folds}

Another promising class of compactifications goes under the name
of  F-theory \cite{VafaFt}. In more mondane terms, F-theory
describes Type IIB configurations with varying dilaton
$\tau=\tau(z)$, where $z$ are the `internal' coordinates. In
particular in the presence of mutually non local 7-branes, the
dilaton varies and undergoes $SL(2,Z)$ monodromy when carried
around the location of the 7-branes.

In a Weierstrass model for the auxiliary torus fibration: $y^2 =
x^3 + f_8(z) x + g_{12}(z) $, with $z$ coordinates on a complex
3-fold base $\cB$, the locations of 7-branes is coded in  the
discriminant locus \be \Delta_{24}(z) = 27 f_8(z)^3 + 4
g^2_{12}(z)= \prod_i (z-z_i) \ee The local behaviour of the
complexified dilaton is: \be2\pi i \tau(z) \approx \log (z-z_i)\ee

In the orientifold limit, in which non-mutually local  7-branes
form a bound state with monodromy $\tau \rightarrow \tau - 4$, the
dilaton is locally constant (up to monodromy) and the rest of the
7-branes are D7-branes \cite{SenFt}.

Recently, there has been a noteworthy revival of local GUT's from
F-theory on CY 4-folds. In local models with 7-branes wrapping Del
Pezzo surfaces  \cite{VafaatStrings09}, it has been shown that one
can turn on `magnetic' $F_Y$ fluxes breaking $SU(5)$ to the SM. A
specific choice of four stacks $\cS, \cP, \cQ, \cR$ seems capable
of producing all kown (chiral) matter at intersections as well as
the necessary Yukawa's at triple intersections, with a concrete
prediction for the Cabibbo angle, or rather of the texture in the
CKM matrix, modulo the assumptions that the local model could be
embedded in a globally consistent one and that the additional
branes and fluxes could stabilize the moduli controlling the
strength of the Yukawa couplings \cite{globFt}.

Notwithstanding its elegance in re-producing  low-energy
phenomenology, F-theory lacks even more than M-theory a concrete
microscopic description. Most of the results are valid only at low
energy in that they are based on complex algebraic properties that
can only determine light or massless states. Massive strings /
branes are completely beyond reach together with all the resulting
`threshold' corrections to gauge and Yukawa couplings.

From now on, we will focus on the perturbative framework of
open and closed unoriented strings.

\section*{Intersecting vs Magnetized Branes}

First of all it is important to observe that magnetized D-branes
support massless chiral fermions \cite{MagnDb}. The chiral
asymmetry is governed by an index theorem \cite{MBET12}
$$ \cI_{ab} = {W_a W_b \over {2\pi}^n n!} \int_{\cM_{2n}} (q_aF_a + q_b
F_b)^n $$ where $W$'s denote the wrapping numbers and $F$'s the
magnetic fields on the two stacks of branes where the two ends of
an open string are located. Magnetized branes are characterized by
the modified boundary conditions \be \de_n X^\mu = 2\pi \ap
F^\mu{}_\nu \de_t X^\nu\ee

Under T -duality $\de_t Y \rightarrow \de_n Y$ yielding  the
boundary conditions for a rotated brane \cite{Bala}. Thus
relatively magnetized branes become intersecting branes under
T-duality. The intersection angles are related to the magnetic
fluxes via
$$\tan(\beta^{ij}_{ab}) = q_aF^{ij}_a + q_b F^{ij}_b $$
The condition for SUSY is
$$ \beta^{12}_{ab} + \beta^{34}_{ab} + \beta^{56}_{ab} = 0$$
More generally, one can consider `co-isotropic' branes \ie
magnetized branes intersecting with other (magnetized) branes.

A peculiar feature of this class of models, \ie vacuum
configurations with Intersecting and/or Magnetized D-Branes, is
the lack of a GUT structure. Since the (tree-level) gauge
couplings \cite{Uberalles}
$$(g_{_{YM}}^{(a)})^2 =
{g_s \over \sqrt{\det(\cG_{(a)} + 2\pi\ap\cF_{(a)})}} $$ depend on
the magnetic flux $\cF_{(a)}$ and metric $\cG_{(a)}$ induced on
the internal cycle wrapped by a given stack of D-branes, GUT would
be quite unnatural or un-necessary.

\section*{Low String Tension and Large Extra Dimensions}

Very large internal volume are compatible with low string tension
($M_s \approx TeV$) and $g_s<<1$ \cite{AAHDD} since
$$M_{Pl} \approx {M_s \sqrt{V_{int}}\over g_s} $$ One arrives at
the idea of un-oriented D-brane worlds, whereby (N-MS)SM lives on
a stack of branes while gravity lives in the 10-d bulk. The
weakness of gravity arises from `dilution' in the extra dimensions
$$ M_{Pl,(4)}^{2} = M_{Pl,(D)}^{ D-2 } L_{int}^{D-4} $$
A slightly alternative scenario relies on `warped'
extra-dimensions \cite{RS12}. In the simplest instance the
universe is a slice of AdS space: gravity is localized on the UV
brane at one end, while (N-MS)SM is localized on the TeV brane at
the other end.

Spectacular signals may be detected at LHC from KK excitations
and/or Regge recurrences! Indeed, many string amplitudes, such as
MHV amplitudes and SUSY related ones, relevant for collisions at
accelerators, contain a simple Veneziano form-factor
\cite{PerelsteinPeskin, Dudasetal, ChialvaIengo, MBAVS, AnchorLST}
$$\cA^{MHV}_{4,st} (s,t,u) = \cA^{MHV}_{4,ft} (s,t,u) {\Gamma(1-\ap s)
\Gamma(1-\ap t) \over \Gamma(1-\ap u)}$$

\section*{Embedding (N-MS)SM in String Theory}

In order to embed the SM in this context, one needs at least four
stacks of intersecting / magnetized branes \cite{Uberalles,
Kirbook}
$$U(3) \times U(2) \times U(1) \times U(1)' \rightarrow SU(3)_c
\times SU(2)_w \times U(1)_Y \times U(1)^3$$ Hypercharge embedding
$Q_Y = a Q_A + b Q_B + c Q_C + d Q_D$ leads to two favourite
choices \cite{YA,YB} and many other possibilities \cite{PAetalY}.

Actually the presence of extra anomalous $U(1)'s$ requires not
only a generalized GS mechanism but also generalized Chern-Simons
couplings \cite{PAMBEDEKetc}.

Some non-anomalous $U(1)$'s become massive anyway due to anomalies
in higher dimensions \cite{PAanom}. Baryon and lepton numbers may
appear as `global' remnants of some (anomalous) gauge symmetries.

\section*{Anomalous U(1)'s and generalized CS couplings}

The need for GCS couplings can be easily argued for. Chiral
anomaly receives a one-loop contribution from the fermion
determinant
$$ \delta \cL_{1-loop} = t_{ijk} \alpha^i F^j \wedge F^k $$
with $ t_{ijk} = \sum_f Q^i_f Q^j_f Q^k_f $ totally symmetric

Closed-string axions in (twisted) R-R sectors couple to
open-string vector bosons and behave as St\"uckelberg fields
$$\cL_{ax} = (d\beta^I - M^I_i A^i)\wedge *(d\beta^I - M^I_i A^i)
+ C^I_{jk} \beta_I F^j \wedge F^k $$ indeed the axionic shifts can
be gauged by (anomalous) vector bosons  $$\delta A^i =
d\alpha^i\qquad \delta \beta^I = M^I_i \alpha^i $$

As a result one finds $$\delta \cL_{ax} = M^I_i \alpha^i C^I_{jk}
\beta_I F^j \wedge F^k$$ with $M^I_i C^I_{jk}$ not totally
symmetric. Introducing generalized CS couplings, however,
$$\cL_{GCS} = E_{ij,k} A^i\wedge A^j \wedge F^k$$ so that
$\delta \cL_{GCS} = E_{ij,k} \alpha^i\wedge F^j \wedge F^k$ leads
to complete anomaly cancellation \cite{PAMBEDEKetc} $$ t_{ijk} -
M^I_k \alpha^i C^I_{ij} = - E_{ij,k} $$

GCS terms can involve the vector boson that gauges hypercharge and
some vector boson $Z'$ that gauges an (anomalous) $U(1)$. After
electro-weak symmetry breaking, this would to a trilinear
couplings of $Z'$ with $Z$ and $\gamma$ with peculiar signals at
LHC for $M_{Z'}$ at the TeV scale.

\section*{Possible experimental signatures}

At LHC one could detect the decay $ Z' \rightarrow Z + \gamma $,
by observing a rather sharp line in the spectrum of $\gamma$. For
instance, in the $Z'$ rest frame one has

$$E_{\gamma} = { M_{Z'}^2 - M_Z^2 \over 2 M_{Z'}} $$

Similarly one can observe other decay channels, such as $ Z'
\rightarrow Z + Z^* $ or $ Z' \rightarrow \gamma + \gamma^* $,
that require one (slightly) off-shell product, otherwise the decay
amplitude would vanish due to Bose statistics for `identical'
vectors.

Even if it were visible at LHC, the above processes would not
necessarily be a signature of String Theory but could be taken as
circumstantial evidence for low-tension / LED strings.

Moreover, in this class of models, the axino, the fermionic
superpartner of the axio-dilaton contributing to anomaly
cancellation, may turn out to be the LSP (rather the `standard'
neutralino) and thus a candidate for dark matter. Analysis of
signals from cosmic rays experiments, such as PAMELA and FERMI,
are under way \cite{AnasFucetal,Coriaetal,Dudas,Kiritetal}.

\section*{Further phenomenological considerations}

  Despite some success in embedding (N-MS)SM in vacuum
  configurations with open and unoriented strings there are some
problems with interactions at the perturbative level

\begin{itemize}
  \item{In $U(5)$ (susy) GUT's
$$
H^d_{{\bf 5}^*_{-1}} F^c_{{\bf 5}^*_{-1}} A_{{\bf 10}_{+2}} \quad
OK \quad {\bf but} \quad H^u_{{\bf 5}_{+1}} A_{{\bf 10}_{+2}}
A_{{\bf 10}_{+2}}  \quad KO
$$
the latter Yukawa is forbidden by (global, anomalous) $U(1)$
invariance, though
 compatible with $SU(5)$: there is no way to generate $\eps^{abcde}$
in perturbation theory from
 Chan-Paton factors}

\item{R-handed (s)neutrino masses $ W_M = M_R NN $ are forbidden
by \eg $U(1)_{B-L}$ in Pati-Salam like models $SO(6)\times SO(4)
\rightarrow SU(3)\times SU(2)_L \times SU(2)_R \times U(1)_{B-L}$}

\item{$\mu$-term in MSSM $ W_{\mu} = \mu H_1 H_2 $ is typically
forbidden by extra (anomalous) $U(1)$'s or their `global'
remnants.}
\end{itemize}

All the above couplings can however be generated by `stringy'
instantons, in particular un-oriented D-brane instantons
\cite{unDbinst}, that result from wrapping Euclidean branes around
non-contractible cycles in the internal space \cite{DSW,BBS,Doug}.

\section*{Two classes of unoriented D-brane instantons}

There are two broad classes of un-oriented D-brane instantons

\begin{itemize}

\item{ `Gauge' instantons correspond to EDp-branes wrapping the
same cycle $C$ as a stack of background D(p+4)-branes, they induce
$F=*_4{F}$ and generate effects with strength
$$ e^{-W_{p+1}(C)/g_s \ell_s^{p+1}}= e^{-1/g_{YM}^2}$$
The resulting system has roughly speaking 4 N-D directions
(spacetime) and allow to make the (super)ADHM construction rather
intuitive!}

\item{ `Exotic' instantons correspond to EDp'-branes wrapping a
cycle $C'$ not wrapped by any stack of background D(p+4)-branes.
In some cases $F\wedge F =*_8{F\wedge F}$ \cite{Billetal8}, the
strength is given by
$$ e^{-W_{p'+1}(C')/g_s \ell_s^{p'+1}} \neq e^{-1/g_{YM}^2}$$
and roughly speaking the system has 8 N-D directions (spacetime +
internal).}

\end{itemize}

\section*{Non-perturbative effects}

For phenomenological purposes it is crucial to identify the sort
of non-perturbative effects induced by the two classes of
un-oriented D-brane instantons.

 `Gauge instantons' may generate
VY-ADS-like superpotentials of the form
$$W = {\Lambda^{\beta} \over \phi^{\beta - 3}} \quad with \quad
\Lambda = M_s e^{-{8\pi^2\over \beta g_{_{YM}}^2(M_s)}}$$ with
$\beta$ the one-loop $\beta$-function coefficient of the (quiver)
gauge theory living on the background stack of D-branes.

`Exotic instantons' may generate non-perturbative superpotentials
of the form
$$W = M_s^{3-n} e^{-S_{EDp'}(C')} \phi^n \quad (n=1, 2,3,...)$$

In both cases, the thumb rule for a non-vanishing result is the
presence of two exact fermionic zero-modes not lifted by (D-term)
interactions.

Explicit computations for intersecting branes and branes at
singularities (quiver gauge theories) have confirmed the above
pattern \cite{Billetal,Blumetal,Ibanetal, MBetal,Kachetal,...}

(Partial) moduli stabilization and SUSY breaking in combination
with (R-R) fluxes and field-dependent FI terms for anomalous
$U(1)$'s can be achieved in some special cases
\cite{MBdsb,Uranga,Francoetal}.

Unoriented D-brane instantons with a different structure of
fermionic zero-modes may lead to threshold corrections to gauge
couplings and other higher-derivative F-terms. One can learn a lot
about multi-instantons or poly-instantons from Heterotic / Type I
duality \cite{BlumSS, MBMora, CDMP, ...}.

\section*{Fluxes, dualities and ... Landscape}

In the intersecting brane picture it is relatively easy to
identify (non) susy flux vacua with $\Lambda \le 0$. Indeed many
near-horizon geometries of this kind of configurations are of the
form $AdS_4 \times G/H$ \cite{LustAdS} and a variant of the
attractor mechanism \cite{FKSetal} is at work. Most of the
would-be moduli are stabilized by the flux-generated scalar
potential. Unfortunately, very much as for $AdS_5\times S^5$, one
has some control of the only in the low-energy supergravity
approximation. The spectrum of massive strings is completely
unknown, as customary in the presence of (R-R) fluxes. Conditions
for decoupling of massive KK excitations can be met in some cases
but deriving their spectrum by means of harmonic analysis on $G/H$
may be rather involved and tedious.

In addition to these technical problems, the crucial issue is the
up-lift to De Sitter with a very small cosmological constant
$0<\Lambda << M$ \cite{KKLT,Large}. All the mechanisms proposed so
far have some drawbacks. Moreover, there seems to be some tension
between moduli stabilization and chirality to the extent that, at
present, there is not a single model enjoying at least two of the
following three properties

\begin{itemize}

\item{Embed N-MSSM with its chiral content}

\item{Stabilize all the moduli}

\item{Break supersymmetry with very small dark energy}

\end{itemize}

Maybe one should not look for specific (classes of) models but
rather extract statistical predictions and correlations among
macroscopic observables in the vast landscape of vacua \eg number
of generations vs rank of the gauge group, Yukawa couplings vs
average gauge couplings, ...

Another possibility that is recently emerging is what may be
termed `holographic phenomenology'. It may well be possible that,
thanks to the holographic principle, String Theory could find some
experimental evidence in previously unimaginable systems such as
\begin{itemize}
\item{Quark-Gluon plasma in HIC \cite{HoloQCD}}
\item{Superconductors \cite{HoloSupCond}} \item{Hydrodynamics
\cite{HoloHydro}} \item{Kerr BH's and Super-stars
\cite{HoloAstro}} \item{Gravity at a Lifshitz point
\cite{HoloHora}}
\end{itemize}

\section*{Outlook}

In conclusion, there are several viable phenomenological scenari
in String Theory but no one is to be particularly preferred since
predictions are still rather vague.

To a certain extent vacuum configurations with open and unoriented
strings that allow for low string tension and large extra
dimensions can be falsified even at LHC. They tend to predict
extra (anomalous) Z', Regge recurrences, and KK excitations at the
TeV scale. In some cases, a non-standard `neutralino', the axino,
may be the LSP and thus the relevant WIMP candidate for dark
matter.

Since a complete global picture is lacking: the Large Scale
Structure of the String Landscape is to be explored at a much
deeper level with an eye wide open onto Holography Phenomenology.

\vskip 1cm \noindent {\large {\bf Acknowledgments}} \vskip 0.2cm I
would like to thank P.~Anastasopoulos, C.~Angelantonj,
E.~Kiritsis, S.~Kovacs, J.~F.~Morales, G.~Pradisi, G.~C.Rossi,
A.~Sagnotti, Ya.~Stanev for long-lasting and very enjoyable
collaboration on various aspects of string phenomenology. I would
like to thank STIAS and NIThep for hospitality during completion
of this project. This work was partially supported by the ERC
Advanced Grant n.226455 {\it ``Superfields''}, by the Italian
MIUR-PRIN contract 20075ATT78 and by the NATO grant
PST.CLG.978785.


\begin{thebibliography}{99}
\bibitem{Candetal}
  P.~Candelas, G.~T.~Horowitz, A.~Strominger and E.~Witten,
  Nucl.\ Phys.\  B {\bf 258} (1985) 46.

\bibitem{HetPhen}
  L.~E.~Ibanez, J.~Mas, H.~P.~Nilles and F.~Quevedo,
  Nucl.\ Phys.\  B {\bf 301}, 157 (1988).

\bibitem{Kirbook}
  E.~Kiritsis,
{\it  Princeton, USA: Univ. Pr. (2007) 588 p}


\bibitem{Aug} 
  A.~Sagnotti,
  arXiv:hep-th/0208020.

\bibitem{CAASrev} 
  For a review see \eg C.~Angelantonj and A.~Sagnotti,
  Phys.\ Rept.\  {\bf 371}, 1 (2002)
  [Erratum-ibid.\  {\bf 376}, 339 (2003)]
  [arXiv:hep-th/0204089].

\bibitem{MBASGP}
  M.~Bianchi and A.~Sagnotti,
  Phys.\ Lett.\  B {\bf 211}, 407 (1988).
  G.~Pradisi and A.~Sagnotti,
  Phys.\ Lett.\  B {\bf 216}, 59 (1989).
  M.~Bianchi and A.~Sagnotti,
  Phys.\ Lett.\  B {\bf 231}, 389 (1989).

\bibitem{MBAS123}
  M.~Bianchi and A.~Sagnotti,
  Phys.\ Lett.\  B {\bf 247}, 517 (1990).
  M.~Bianchi and A.~Sagnotti,
  Nucl.\ Phys.\  B {\bf 361}, 519 (1991).
  C.~Angelantonj, M.~Bianchi, G.~Pradisi, A.~Sagnotti and Y.~S.~Stanev,
  Phys.\ Lett.\  B {\bf 387}, 743 (1996)
  [arXiv:hep-th/9607229].

\bibitem{GSSmech}
  A.~Sagnotti,
  Phys.\ Lett.\  B {\bf 294} (1992) 196
  [arXiv:hep-th/9210127].

\bibitem{BPSt}
  M.~Bianchi, G.~Pradisi and A.~Sagnotti,
  Nucl.\ Phys.\  B {\bf 376}, 365 (1992).


\bibitem{ABPSS} 
  C.~Angelantonj, M.~Bianchi, G.~Pradisi, A.~Sagnotti and Y.~S.~Stanev,
  Phys.\ Lett.\  B {\bf 385}, 96 (1996)
  [arXiv:hep-th/9606169].


\bibitem{MagnDb}
  M.~Bianchi and Y.~S.~Stanev,
  Nucl.\ Phys.\  B {\bf 523} (1998) 193
  [arXiv:hep-th/9711069].
  C.~Angelantonj, I.~Antoniadis, G.~D'Appollonio, E.~Dudas and A.~Sagnotti,
  Nucl.\ Phys.\  B {\bf 572}, 36 (2000)
  [arXiv:hep-th/9911081].

\bibitem{BCFT}
  M.~Bianchi, G.~Pradisi and A.~Sagnotti,
  Phys.\ Lett.\  B {\bf 273}, 389 (1991).
 G.~Pradisi, A.~Sagnotti and Y.~S.~Stanev,
  Phys.\ Lett.\  B {\bf 381}, 97 (1996)
  [arXiv:hep-th/9603097].

\bibitem{DaiLP}
  J.~Dai, R.~G.~Leigh and J.~Polchinski,
  Mod.\ Phys.\ Lett.\  A {\bf 4}, 2073 (1989).

\bibitem{JoeP}
  J.~Polchinski,
  Phys.\ Rev.\ Lett.\  {\bf 75}, 4724 (1995)
  [arXiv:hep-th/9510017].


\bibitem{HullTownDemo}
  P.~K.~Townsend,
  arXiv:hep-th/9507048.


\bibitem{NS5} 
  A.~Dabholkar, J.~P.~Gauntlett, J.~A.~Harvey and D.~Waldram,
  Nucl.\ Phys.\  B {\bf 474}, 85 (1996)
  [arXiv:hep-th/9511053].
  D.~S.~Berman,
  Phys.\ Rept.\  {\bf 456}, 89 (2008)
  [arXiv:0710.1707 [hep-th]].


\bibitem{WitDb}
  E.~Witten,
  Nucl.\ Phys.\  B {\bf 460}, 335 (1996)
  [arXiv:hep-th/9510135].


\bibitem{HullTownU}
  C.~M.~Hull and P.~K.~Townsend,
  Nucl.\ Phys.\  B {\bf 438}, 109 (1995)
  [arXiv:hep-th/9410167].

\bibitem{WitDynVD}
  E.~Witten,
  Nucl.\ Phys.\  B {\bf 443}, 85 (1995)
  [arXiv:hep-th/9503124].



\bibitem{PolWit}
J.~Polchinski and E.~Witten,
  Nucl.\ Phys.\  B {\bf 460}, 525 (1996)
  [arXiv:hep-th/9510169].



\bibitem{HullHet}
  C.~M.~Hull,
  Phys.\ Lett.\  B {\bf 357}, 545 (1995)
  [arXiv:hep-th/9506194].

\bibitem{Dabhet}
  A.~Dabholkar,
  Phys.\ Lett.\  B {\bf 357}, 307 (1995)
  [arXiv:hep-th/9506160].

\bibitem{IHtests}
  C.~Bachas, C.~Fabre, E.~Kiritsis, N.~A.~Obers and P.~Vanhove,
  Nucl.\ Phys.\  B {\bf 509}, 33 (1998)
  [arXiv:hep-th/9707126].
  E.~Gava, J.~F.~Morales, K.~S.~Narain and G.~Thompson,
  Nucl.\ Phys.\  B {\bf 528}, 95 (1998)
  [arXiv:hep-th/9801128].
  M.~Bianchi, E.~Gava, F.~Morales and K.~S.~Narain,
  Nucl.\ Phys.\  B {\bf 547}, 96 (1999)
  [arXiv:hep-th/9811013].


\bibitem{nonBPS}
  M.~Frau, L.~Gallot, A.~Lerda and P.~Strigazzi,
  Nucl.\ Phys.\  B {\bf 564}, 60 (2000)
  [arXiv:hep-th/9903123].



\bibitem{HetIIAK3}
  E.~Kiritsis, N.~A.~Obers and B.~Pioline,
  JHEP {\bf 0001}, 029 (2000)
  [arXiv:hep-th/0001083].

\bibitem{VafaFt}
  C.~Vafa,
  Nucl.\ Phys.\  B {\bf 469}, 403 (1996)
  [arXiv:hep-th/9602022].

\bibitem{JMholoetal}
  J.~M.~Maldacena,
  Adv.\ Theor.\ Math.\ Phys.\  {\bf 2}, 231 (1998)
  [Int.\ J.\ Theor.\ Phys.\  {\bf 38}, 1113 (1999)]
  [arXiv:hep-th/9711200].
  E.~Witten,
  Adv.\ Theor.\ Math.\ Phys.\  {\bf 2}, 253 (1998)
  [arXiv:hep-th/9802150].
  S.~S.~Gubser, I.~R.~Klebanov and A.~M.~Polyakov,
  Phys.\ Lett.\  B {\bf 428}, 105 (1998)
  [arXiv:hep-th/9802109].


\bibitem{RS12}
  L.~Randall and R.~Sundrum,
  Phys.\ Rev.\ Lett.\  {\bf 83}, 4690 (1999)
  [arXiv:hep-th/9906064].
  L.~Randall and R.~Sundrum,
  Phys.\ Rev.\ Lett.\  {\bf 83}, 3370 (1999)
  [arXiv:hep-ph/9905221].



\bibitem{HoloPhen} 
  U.~Gursoy, E.~Kiritsis, L.~Mazzanti and F.~Nitti,
  Nucl.\ Phys.\  B {\bf 820}, 148 (2009)
  [arXiv:0903.2859 [hep-th]].


\bibitem{GKP} 
  S.~B.~Giddings, S.~Kachru and J.~Polchinski,
  Phys.\ Rev.\  D {\bf 66}, 106006 (2002)
  [arXiv:hep-th/0105097].


\bibitem{FluxGaugSUGRA}
  B.~de Wit, H.~Nicolai and H.~Samtleben,
  JHEP {\bf 0802}, 044 (2008)
  [arXiv:0801.1294 [hep-th]].

\bibitem{unDbinst} 
For a recent review see \eg  R.~Blumenhagen, M.~Cvetic, S.~Kachru
and T.~Weigand,
  arXiv:0902.3251 [hep-th].

\bibitem{GSW}
  M.~B.~Green, J.~H.~Schwarz and E.~Witten,
{\it  Cambridge, Uk: Univ. Pr. ( 1987) 596 P. ( Cambridge
Monographs On Mathematical Physics)}



\bibitem{racetrack} 
  L.~J.~Dixon,

\bibitem{HetTors}
  A.~Strominger,
  Nucl.\ Phys.\  B {\bf 274}, 253 (1986).



\bibitem{HullDoubling}
  C.~M.~Hull and R.~A.~Reid-Edwards,
  arXiv:hep-th/0503114.



\bibitem{MBLRasymmDb}
  M.~Bianchi,
  Nucl.\ Phys.\  B {\bf 805}, 168 (2008)
  [arXiv:0805.3276 [hep-th]].






\bibitem{Uberalles}
  R.~Blumenhagen, B.~Kors, D.~Lust and S.~Stieberger,
  Phys.\ Rept.\  {\bf 445} (2007) 1
  [arXiv:hep-th/0610327].
  A.~M.~Uranga,
  JHEP {\bf 0901}, 048 (2009)
  [arXiv:0808.2918 [hep-th]].


\bibitem{MBJFM}
  M.~Bianchi and J.~F.~Morales,
  JHEP {\bf 0003}, 030 (2000)
  [arXiv:hep-th/0002149].



\bibitem{AIUM}
  G.~Aldazabal, D.~Badagnani, L.~E.~Ibanez and A.~M.~Uranga,
  JHEP {\bf 9906}, 031 (1999)
  [arXiv:hep-th/9904071].


\bibitem{SenFt}
  A.~Sen,
  Phys.\ Rev.\  D {\bf 55}, 7345 (1997)
  [arXiv:hep-th/9702165].

\bibitem{PureSpin}
  N.~Berkovits,
  arXiv:hep-th/0209059.


\bibitem{Hybrid}
  N.~Berkovits and B.~C.~Vallilo,
  Nucl.\ Phys.\  B {\bf 624}, 45 (2002)
  [arXiv:hep-th/0110168].


\bibitem{superWZW}
  C.~Candu, V.~Mitev, T.~Quella, H.~Saleur and V.~Schomerus,
  arXiv:0908.0878 [hep-th].


\bibitem{BMNetal}
  D.~E.~Berenstein, J.~M.~Maldacena and H.~S.~Nastase,
  AIP Conf.\ Proc.\  {\bf 646}, 3 (2003).

\bibitem{BeisRev}
  G.~Arutyunov and S.~Frolov,
  J.\ Phys.\ A  {\bf 42}, 254003 (2009)
  [arXiv:0901.4937 [hep-th]].


\bibitem{KonNoi}
  M.~Bianchi, S.~Kovacs, G.~Rossi and Y.~S.~Stanev,
  JHEP {\bf 0105}, 042 (2001)
  [arXiv:hep-th/0104016].

\bibitem{Vieiraetal}
  N.~Gromov, V.~Kazakov and P.~Vieira,
  arXiv:0906.4240 [hep-th].


\bibitem{AchaWit}
  B.~S.~Acharya and E.~Witten,
  arXiv:hep-th/0109152.

\bibitem{AchaKaneKumar}
  B.~S.~Acharya, K.~Bobkov, G.~L.~Kane, J.~Shao and P.~Kumar,
  Phys.\ Rev.\  D {\bf 78}, 065038 (2008)
  [arXiv:0801.0478 [hep-ph]].

\bibitem{M2new}
  J.~Bagger and N.~Lambert,
  Phys.\ Rev.\  D {\bf 75}, 045020 (2007)
  [arXiv:hep-th/0611108].
  A.~Gustavsson,
  Nucl.\ Phys.\  B {\bf 811}, 66 (2009)
  [arXiv:0709.1260 [hep-th]].
  O.~Aharony, O.~Bergman, D.~L.~Jafferis and J.~Maldacena,
  JHEP {\bf 0810}, 091 (2008)
  [arXiv:0806.1218 [hep-th]].


\bibitem{VafaatStrings09}
  J.~J.~Heckman, A.~Tavanfar and C.~Vafa,
  arXiv:0906.0581 [hep-th].


\bibitem{globFt}
  R.~Blumenhagen, T.~W.~Grimm, B.~Jurke and T.~Weigand,
  arXiv:0908.1784 [hep-th].

\bibitem{MBET12}
  M.~Bianchi and E.~Trevigne,
  JHEP {\bf 0601}, 092 (2006)
  [arXiv:hep-th/0506080].
  M.~Bianchi and E.~Trevigne,
  JHEP {\bf 0508}, 034 (2005)
  [arXiv:hep-th/0502147].




\bibitem{Bala}
  V.~Balasubramanian and R.~G.~Leigh,
  Phys.\ Rev.\  D {\bf 55}, 6415 (1997)
  [arXiv:hep-th/9611165].

\bibitem{AAHDD}
  I.~Antoniadis, N.~Arkani-Hamed, S.~Dimopoulos and G.~R.~Dvali,
  Phys.\ Lett.\  B {\bf 436}, 257 (1998)
  [arXiv:hep-ph/9804398].

\bibitem{PerelsteinPeskin}
  S.~Cullen, M.~Perelstein and M.~E.~Peskin,
  Phys.\ Rev.\  D {\bf 62}, 055012 (2000)
  [arXiv:hep-ph/0001166].

\bibitem{Dudasetal}
 E.~Dudas and C.~Timirgaziu,
  Nucl.\ Phys.\  B {\bf 716}, 65 (2005)
  [arXiv:hep-th/0502085].

\bibitem{ChialvaIengo}
  D.~Chialva, R.~Iengo and J.~G.~Russo,
  Phys.\ Rev.\  D {\bf 71}, 106009 (2005)
  [arXiv:hep-ph/0503125].

\bibitem{MBAVS}
  M.~Bianchi and A.~V.~Santini,
  JHEP {\bf 0612}, 010 (2006)
  [arXiv:hep-th/0607224].

\bibitem{AnchorLST}
  L.~A.~Anchordoqui, H.~Goldberg, D.~Lust, S.~Nawata, S.~Stieberger and T.~R.~Taylor,
  Nucl.\ Phys.\  B {\bf 821}, 181 (2009)
  [arXiv:0904.3547 [hep-ph]].


\bibitem{YA} 
  I.~Antoniadis, E.~Kiritsis, J.~Rizos and T.~N.~Tomaras,
  Nucl.\ Phys.\  B {\bf 660}, 81 (2003)
  [arXiv:hep-th/0210263].


\bibitem{YB} 
  G.~Aldazabal, S.~Franco, L.~E.~Ibanez, R.~Rabadan and A.~M.~Uranga,
  JHEP {\bf 0102}, 047 (2001)
  [arXiv:hep-ph/0011132].


\bibitem{PAetalY}
  P.~Anastasopoulos, T.~P.~T.~Dijkstra, E.~Kiritsis and A.~N.~Schellekens,
  Nucl.\ Phys.\  B {\bf 759}, 83 (2006)
  [arXiv:hep-th/0605226].


\bibitem{PAMBEDEKetc}
  P.~Anastasopoulos, M.~Bianchi, E.~Dudas and E.~Kiritsis,
  JHEP {\bf 0611}, 057 (2006)
  [arXiv:hep-th/0605225].


\bibitem{PAanom}
  P.~Anastasopoulos,
  JHEP {\bf 0308}, 005 (2003)
  [arXiv:hep-th/0306042].


\bibitem{AnasFucetal}
  P.~Anastasopoulos, F.~Fucito, A.~Lionetto, G.~Pradisi, A.~Racioppi and Y.~S.~Stanev,
  Phys.\ Rev.\  D {\bf 78}, 085014 (2008)
  [arXiv:0804.1156 [hep-th]].


\bibitem{Coriaetal}
  R.~Armillis, C.~Coriano', M.~Guzzi and S.~Morelli,
  Nucl.\ Phys.\  B {\bf 814}, 15679 (2009)
  [arXiv:0809.3772 [hep-ph]].

\bibitem{Dudas}
  E.~Dudas, Y.~Mambrini, S.~Pokorski, A.~Romagnoni and M.~Trapletti,
  JHEP {\bf 0903}, 011 (2009)
  [arXiv:0809.5064 [hep-th]].

\bibitem{Kiritetal} 
  P.~Anastasopoulos, E.~Kiritsis and A.~Lionetto,
  JHEP {\bf 0908}, 026 (2009)
  [arXiv:0905.3044 [hep-th]].


\bibitem{DSW}
  M.~Dine, N.~Seiberg, X.~G.~Wen and E.~Witten,
  Nucl.\ Phys.\  B {\bf 278}, 769 (1986).


\bibitem{BBS}
  K.~Becker, M.~Becker and A.~Strominger,
  Nucl.\ Phys.\  B {\bf 456}, 130 (1995)
  [arXiv:hep-th/9507158].


\bibitem{Doug}
  E.~Witten,
  Nucl.\ Phys.\  B {\bf 460}, 541 (1996)
  [arXiv:hep-th/9511030].


 \bibitem{Billetal8}
  M.~Billo, L.~Ferro, M.~Frau, L.~Gallot, A.~Lerda and I.~Pesando,
  JHEP {\bf 0907}, 092 (2009)
  [arXiv:0905.4586 [hep-th]].


 \bibitem{Billetal}
  M.~Billo, M.~Frau, I.~Pesando, F.~Fucito, A.~Lerda and A.~Liccardo,
  JHEP {\bf 0302}, 045 (2003)
  [arXiv:hep-th/0211250].
  M.~Billo', L.~Ferro, M.~Frau, F.~Fucito, A.~Lerda and J.~F.~Morales,
  JHEP {\bf 0810}, 112 (2008)
  [arXiv:0807.1666 [hep-th]].




 \bibitem{Blumetal}
  R.~Blumenhagen, M.~Cvetic and T.~Weigand,
  Nucl.\ Phys.\  B {\bf 771}, 113 (2007)
  [arXiv:hep-th/0609191].
  M.~Cvetic, R.~2.~Richter and T.~Weigand,
  AIP Conf.\ Proc.\  {\bf 957}, 30 (2007).



 \bibitem{Ibanetal}
  L.~E.~Ibanez and A.~M.~Uranga,
  JHEP {\bf 0703}, 052 (2007)
  [arXiv:hep-th/0609213].
  L.~E.~Ibanez, A.~N.~Schellekens and A.~M.~Uranga,
  JHEP {\bf 0706}, 011 (2007)
  [arXiv:0704.1079 [hep-th]].

 \bibitem{MBetal}
  M.~Bianchi and E.~Kiritsis,
  Nucl.\ Phys.\  B {\bf 782}, 26 (2007)
  [arXiv:hep-th/0702015].
  M.~Bianchi, F.~Fucito and J.~F.~Morales,
  JHEP {\bf 0707}, 038 (2007)
  [arXiv:0704.0784 [hep-th]].



 \bibitem{Kachetal}
  B.~Florea, S.~Kachru, J.~McGreevy and N.~Saulina,
  JHEP {\bf 0705}, 024 (2007)
  [arXiv:hep-th/0610003].


 \bibitem{MBdsb}
  M.~Bianchi, F.~Fucito and J.~F.~Morales,
  JHEP {\bf 0908}, 040 (2009)
  [arXiv:0904.2156 [hep-th]].

 \bibitem{Uranga}
  I.~Garcia-Etxebarria, F.~Saad and A.~M.~Uranga,
  JHEP {\bf 0705}, 047 (2007)
  [arXiv:0704.0166 [hep-th]].

 \bibitem{Francoetal}
  R.~Argurio, M.~Bertolini, S.~F.~Franco and S.~Kachru,
  Fortsch.\ Phys.\  {\bf 55}, 644 (2007).

 \bibitem{BlumSS}
  R.~Blumenhagen and M.~Schmidt-Sommerfeld,
  JHEP {\bf 0807}, 027 (2008)
  [arXiv:0803.1562 [hep-th]].


 \bibitem{MBMora}
  M.~Bianchi and J.~F.~Morales,
  JHEP {\bf 0802}, 073 (2008)
  [arXiv:0712.1895 [hep-th]].


 \bibitem{CDMP}
  P.~G.~Camara, E.~Dudas, T.~Maillard and G.~Pradisi,
  Nucl.\ Phys.\  B {\bf 795}, 453 (2008)
  [arXiv:0710.3080 [hep-th]].


 \bibitem{LustAdS}
  C.~Caviezel, P.~Koerber, S.~Kors, D.~Lust, D.~Tsimpis and M.~Zagermann,
  Class.\ Quant.\ Grav.\  {\bf 26}, 025014 (2009)
  [arXiv:0806.3458 [hep-th]].


 \bibitem{FKSetal}
  S.~Ferrara, R.~Kallosh and A.~Strominger,
  Phys.\ Rev.\  D {\bf 52}, 5412 (1995)
  [arXiv:hep-th/9508072].

 \bibitem{KKLT}
  S.~Kachru, R.~Kallosh, A.~Linde and S.~P.~Trivedi,
  Phys.\ Rev.\  D {\bf 68}, 046005 (2003)
  [arXiv:hep-th/0301240].


 \bibitem{Large}
  J.~P.~Conlon, C.~H.~Kom, K.~Suruliz, B.~C.~Allanach and F.~Quevedo,
  JHEP {\bf 0708}, 061 (2007)
  [arXiv:0704.3403 [hep-ph]].


 \bibitem{HoloQCD}
  S.~S.~Gubser and A.~Karch,
  arXiv:0901.0935 [hep-th].


 \bibitem{HoloSupCond}
  S.~A.~Hartnoll,
  arXiv:0903.3246 [hep-th].


\bibitem{HoloHydro}
  S.~Bhattacharyya, S.~Minwalla and S.~R.~Wadia,
  JHEP {\bf 0908}, 059 (2009)
  [arXiv:0810.1545 [hep-th]].



\bibitem{HoloAstro}
  T.~Hartman, W.~Song and A.~Strominger,
  arXiv:0908.3909 [hep-th].
  J.~de Boer, K.~Papadodimas and E.~Verlinde,
  arXiv:0907.2695 [hep-th].



 \bibitem{HoloHora}
  P.~Horava,
  Phys.\ Rev.\  D {\bf 79}, 084008 (2009)
  [arXiv:0901.3775 [hep-th]].
\end{thebibliography}
\end{document}